\documentclass[]{article}
\usepackage{lmodern}
\usepackage{amssymb,amsmath}
\usepackage{ifxetex,ifluatex}
\usepackage{fixltx2e} % provides \textsubscript
\ifnum 0\ifxetex 1\fi\ifluatex 1\fi=0 % if pdftex
  \usepackage[T1]{fontenc}
  \usepackage[utf8]{inputenc}
\else % if luatex or xelatex
  \ifxetex
    \usepackage{mathspec}
  \else
    \usepackage{fontspec}
  \fi
  \defaultfontfeatures{Ligatures=TeX,Scale=MatchLowercase}
\fi
\IfFileExists{upquote.sty}{\usepackage{upquote}}{}
\IfFileExists{microtype.sty}{%
\usepackage{microtype}
\UseMicrotypeSet[protrusion]{basicmath} % disable protrusion for tt fonts
}{}
\usepackage{hyperref}
\hypersetup{unicode=true,
            pdftitle={A common misapplication of statistical inference: nuisance control with null-hypothesis significance tests},
            pdfauthor={Jona Sassenhagen; Phillip M. Alday},
            pdfkeywords={design, experimental balance, confound, mixed models},
            pdfborder={0 0 0},
            breaklinks=true}
\IfFileExists{parskip.sty}{%
\usepackage{parskip}
}{% else
\setlength{\parindent}{0pt}
\setlength{\parskip}{6pt plus 2pt minus 1pt}
}
\providecommand{\tightlist}{%
  \setlength{\itemsep}{0pt}\setlength{\parskip}{0pt}}
\ifx\paragraph\undefined\else
\let\oldparagraph\paragraph
\renewcommand{\paragraph}[1]{\oldparagraph{#1}\mbox{}}
\fi
\ifx\subparagraph\undefined\else
\let\oldsubparagraph\subparagraph
\renewcommand{\subparagraph}[1]{\oldsubparagraph{#1}\mbox{}}
\fi

\title{A common misapplication of statistical inference: nuisance control with
null-hypothesis significance tests}
\author{Jona Sassenhagen \and Phillip M. Alday}
\date{May 2016}

\begin{document}
\maketitle
\begin{abstract}
Experimental research on behavior and cognition frequently rests on
stimulus or subject selection where not all characteristics can be fully
controlled, even when attempting strict matching. For example, when
contrasting patients to controls, variables such as intelligence or
socioeconomic status are often correlated with patient status.
Similarly, when presenting word stimuli, variables such as word
frequency are often correlated with primary variables of interest. One
procedure very commonly employed to control for such nuisance effects is
conducting inferential tests on confounding stimulus or subject
characteristics. For example, if word length is not \emph{significantly}
different for two stimulus sets, they are considered as matched for word
length. Such a test has high error rates and is conceptually misguided.
It reflects a common misunderstanding of statistical tests: interpreting
significance not to refer to inference about a particular population
parameter, but about 1. the sample in question, 2. the \emph{practical
relevance} of a sample difference (so that a nonsignificant test is
taken to indicate evidence for the absence of relevant differences). We
show inferential testing for assessing nuisance effects to be
inappropriate both pragmatically and philosophically, present a survey
showing its high prevalence, and briefly discuss an alternative in the
form of regression including nuisance variables.
\end{abstract}

\section{Introduction}\label{introduction}

Methods sections in many issues of \emph{Brain \& Language} and similar
journals feature sentences such as

\begin{quote}
Animate and inanimate words chosen as stimulus materials did not differ
in word frequency (\(p\) \textgreater{} 0.05).
\end{quote}

\begin{quote}
Controls and aphasics did not differ in age (\(p\) \textgreater{} 0.05).
\end{quote}

In the following, we discuss the inappropriateness of this practice. A
common problem in brain and behavioral research, where the experimenter
cannot freely determine every stimulus and participant characteristic,
is the control of confounding/nuisance variables. This is especially
common in studies of language. Typically, word stimuli cannot be
constructed out of whole cloth, but must be chosen from existing words
(which differ in many aspects). Stimuli are processed by subjects in the
context of a rich vocabulary; and subject populations have usually been
exposed to very diverse environments and events in their acquisition of
language. A similar problem exists, for example, when comparing controls
to specific populations, such as bilingual individuals or slow readers.
The basic problem researchers are faced with is then to prevent
reporting e.g.~an effect of word length, or bilingualism, when the
effect truly stems from differences in word frequency, or socioeconomic
status, which may be correlated with the variable of interest. A
prevalent method we find in the literature, namely inferential null
hypothesis significance testing (NHST) of stimuli, fails to perform the
necessary control.

\subsection{NHST and nuisance control}\label{nhst-and-nuisance-control}

Often, researchers will attempt to demonstrate that stimuli or
participants are selected so as to concentrate their differences on the
variable of interest, i.e.~reduce confounds, by conducting
null-hypothesis testing such as \(t\)-tests or ANOVA on the potential
confound in addition to or even instead of showing descriptive
statistics in the form of measures of location and scale. The underlying
intuition is that these tests establish whether two conditions differ in
a given aspect and serve as proof that the conditions are ``equal'' on
it. This is, in turn, based on the related, but also incorrect intuition
that significance in NHST establishes that a contrast shows a
\emph{meaningful effect}, and the related issue that non-significant
tests indicate the absence of meaningful effects.

In practice, we find insignificant tests are used as a necessary (and
often sufficient) condition for accepting a stimulus set as
``controlled''. This approach fails on multiple levels.

\begin{itemize}
\item
  Philosophically, these tests are inferential tests being performed on
  closed populations, not random samples of larger populations.
  Statistical testing attempts to make inferences about the larger
  population based on randomly selected samples. Here, the ``samples''
  are not taken randomly, and we are not interested in the population
  they are drawn from, but in the stimuli or subjects themselves. For
  example, in a study on the effects of animacy in language processing,
  we do not care whether the class of animate nouns in the language is
  on average more frequent than the class of inanimate nouns. Instead,
  we care whether the selection of animate nouns \emph{in our stimuli}
  are on average more frequent than the selection of inanimate nouns
  \emph{in our stimuli}. But inferential tests answer the former
  question, not the latter. Tests refer to the population of stimuli
  that will largely not be used, or the population of subjects that will
  not be investigated in the study.
\item
  Pragmatically, beyond being inappropriate, this procedure does not
  test a hypothesis of interest. This procedure tests the null
  hypothesis of ``the populations that these stimuli were sampled from
  do not differ in this feature'', but what we are actually interested
  in is ``the differences in this feature between conditions is not
  responsible for any observed effects''. In other words, this procedure
  tests whether the conditions differ in a certain respect to a
  measurable degree, but not whether that difference actually has any
  meaningful influence on the result.
\item
  Additionally, these tests carry all the usual problems of Null
  Hypothesis Significance Testing (cf. Cohen 1992), including its
  inability to ``accept'' the null hypothesis directly. This means that
  even if the conditions do not differ significantly, we cannot accept
  the hypothesis that they do not differ; we can only say that there is
  not enough evidence to exclude this hypothesis (which we are not
  actually interested in). In typical contexts (e.g.~setting the Type I
  rate to the conventional 5\% level), the power to reject the null
  hypothesis of no differences is low (Button et al. 2013) due to a
  small number of items, meaning that even comparatively large
  differences may be undetected, while in larger sets, even trivially
  small differences may be rejected. Especially with small samples
  (e.g., 10 subjects per group, or 20 items per condition), the
  probability of detecting moderate confound effects is thus low -- even
  if there are substantial differences, tests will not reject the null
  hypothesis, and stimulus sets might be accepted as being balanced
  based on a test with a low probability of rejecting even moderately
  imbalanced samples of such a size.
\end{itemize}

In other words, these tests are incapable of actually informing us about
the influence of potential confounds, but may give researchers a false
sense of security. This inferential stage offers no benefit beyond
examining the descriptive measures of location and scale (e.g.~mean and
standard deviation) and determining if the stimuli groups are ``similar
enough''. For perceptual experiments, there may even be established
discrimination thresholds below which the differences are considered
indistinguishable. A preferred approach is directly examining to what
extent these potential confounds have an influence on the results, such
as by including these confounds in the statistical model. This is often
readily implemented via multiple regression, particularly
``mixed-effect'' approaches (Gelman and Hill 2006; Fox 2016).

\subsection{Randomization checks in clinical
research}\label{randomization-checks-in-clinical-research}

In the context of baseline differences between treatment and control
groups in clinical trials, a similar debate has been waged (e.g. Senn
1994) under the term ``randomization check'' as it refers to checking if
assignment of subjects to treatments has truly been performed randomly.
In interventional clinical trials, assignment can indeed be truly random
(unlike in the kind of study in brain and behavioral sciences we are
referring to here). Yet even here, inferential tests have been judged
inappropriate for achieving their intended aims. Nonetheless, the
clinical trial literature provides important considerations for
experimental design choices, e.g.~the proper way of blocking and
matching (Imai, King, and Stuart 2008), and can thus inform preparing
stimulus sets or participant groups even for non-clinical experiments.

\section{Prevalence}\label{prevalence}

We performed a literature survey of neurolinguistic studies to estimate
the prevalence of inferential tests of nuisance variables (see below for
further details).

\subsection{Qualitative impressions}\label{qualitative-impressions}

Instances of the error can be easily found not only in the literature,
such as this example from the 1980s:

\begin{quote}
the two prime categories were equivalent in text frequency
{[}\ldots{}{]}, and in length (both \(t\)'s \textless{} 1.1)
\end{quote}

Here, the authors deduce equivalence (acceptance of the null) from a
failed test (i.e.~a test where the null cannot be rejected), with
regards to the population of stimuli they did not present rather than
the sample at hand. To estimate how common the problem is in
neurolinguistics, a high-quality neurolinguistic journal, \emph{Brain \&
Language}, was investigated.

\subsection{\texorpdfstring{Quantitative prevalence of the problem in
recent issues of \emph{Brain \&
Language}}{Quantitative prevalence of the problem in recent issues of Brain \& Language}}\label{quantitative-prevalence-of-the-problem-in-recent-issues-of-brain-language}

In total, 86 articles were found where researchers reported known
quantities (e.g.~perfectly measurable characteristics of a fixed set of
stimuli) in their stimulus/materials section, and 58 (\textbf{67\%}) of
these reported inferential statistics of these known values. Of these,
47 (\textbf{81\%}) ``accepted'' the null hypothesis (i.e., implicitly
assumed that stimuli or subjects were matched following a nonsignificant
test). We conclude that in a large fraction of those cases, where
researchers published in \emph{B\&L} are concerned about confounds of
subject groups or experimental stimuli, they conduct inappropriate tests
and misinterpret the results of these tests in a potentially misleading
manner.

Representative statements from every study committing an error as well
as further details on the precise survey methodology are available
online at https://github.com/jona-sassenhagen/statfail.

\section{Simulation}\label{simulation}

We performed a simulation to investigate the impact of inferential tests
of confounding variables. In particular, we find that when the
correlation between the confounding covariate and the outcome measure is
not perfect, testing covariates (instead of their effect on the outcome
variable) can lead to unnecessary rejections of manipulations as
``confounded'' in 50\% or more of studies for even large
effects.\footnote{20 items each for 2 groups, Cohen's \(d = 2\) for
  manipulation, Pearson's \(r = 0.75\) between feature and dependent
  variable; rejection rate \textgreater{} 60\% with Cohen's \(d = 1\)
  for covariate between groups, rejection rate = 75\% with Cohen's
  \(d = 2\) for covariate between groups. See the
  \href{http://rpubs.com/palday/statfail}{full output on RPubs
  (http://rpubs.com/palday/statfail)} for more information.}

The results of this simulation for various settings (e.g.~effect size,
confound size, etc.) are available online on
\href{http://rpubs.com/palday/statfail}{RPubs
(http://rpubs.com/palday/statfail)}, while an interactive version is
available online at
\href{https://palday.shinyapps.io/statfail/}{ShinyApps
(https://palday.shinyapps.io/statfail/)}. All source code (in R) is
available via Zenodo (DOI: 10.5281/zenodo.58750), including the ability
to run the simulation on a local computer.

\section{Discussion and
recommendation}\label{discussion-and-recommendation}

In sum, NHST control of nuisance variables is prevalent and
inappropriate, based on a flawed application of statistics to an
irrelevant hypothesis. Proper nuisance control (of known and measurable
variables) is not complex, although it can require more effort and
computer time.

Researchers should still use descriptive statistics to demonstrate the
success of balancing. That is, quantifying e.g.~differences between
stimuli via variances, raw means and standardized means (Cohen's
\emph{d}), and correlation coefficients, which many researchers already
often do, can be highly informative, and should be routinely done. For
more complex designs, cross-correlation matrices can visualize the
degree of confounding. In contrast, \(p\) values from statistical tests
on the stimulus properties offer no reliable, objective guideline.

To directly and objectively estimate the influence of a set of stimuli
on the dependent variables of interest, researchers can include
confounds in their statistical model for the data. For traditional
\(t\)-tests, ANOVAs and regression models, this corresponds to using
multiple regression with the confounds as additional nuisance factors
(including continuous factors). In multiple regression, all parameters
are jointly estimated, and assuming the assumptions of the linear model
are fulfilled (including all relevant variables being present and
homoskedasticity of errors) and the included variables are reliably
measured (Westfall and Yarkoni 2016), these estimates are unbiased (in
the sense of a Best Linear Unbiased Estimator). Thus, a manipulation
effect estimated by a model also containing nuisance variables
corresponds to the effect of manipulation while accounting for nuisance
influence. Importantly, to prevent \(p\)-value ``fishing'', the choice
of selecting covariates to include must be made on principled grounds,
either \emph{a priori} or via unbiased model selection procedures.

Hierarchical/multilevel modeling (a.k.a mixed-effects modeling; see also
Pinheiro and Bates 2000; Gelman and Hill 2006; Fox 2016) provides the
necessary extension to the regression procedure for repeated-measures
designs. Multilevel regression models (computed with e.g \emph{lme4}
(Bates et al. 2015)) have the additional advantage of accounting for the
combined variance of subjects and items in one model (Clark 1973;
Baayen, Davidson, and Bates 2008; Judd, Westfall, and Kenny 2012) and
automatically provide a summary of correlation between effects.

One problem in this context is that these stimulus confounds are often
correlated with one another, the dependent variables, and the
independent variables of interest (e.g., word frequency and word length
correlate). Under multicollinearity, standard errors may be inflated.
The main technique for dealing with collinearity is one that researchers
traditionally already employ: attempting to balance stimulus/subject
selection so that differences in confounds are minimised, e.g.~via
matching or blocking. That is, matching should generally still be
performed in addition to multivariate estimation.

Finally, effective parameter estimation in complex regression models
requires more data, as power is lost with each additional parameter
being estimated. We view this as a good thing because studies in the
brain and behavioral sciences are chronically underpowered (Button et
al. 2013).

Thus, our recommendations for nuisance control are:

\begin{itemize}
\tightlist
\item
  attempt to match nuisance variables to a reasonable degree
\item
  use descriptive, but not inferential statistics to guide stimulus
  selection
\item
  add potentially confounding variables as covariates into the final
  data analysis process
\item
  use larger samples to provide adequate power
\end{itemize}

Each step in this list is (hopefully) uncontroversial and helpful,
unlike null-hypothesis testing of stimulus balance.

\section{Acknowledgements}\label{acknowledgements}

We thank Sarah Tune for helpful discussion and Tal Linzen for bringing
to our attention the randomization check literature; Katja Starikova,
Miriam Burk and Antonia Götz are to be thanked for collecting the survey
data. This work was supported in part by the German Research Foundation
(BO 2471/3-2) and by the ERC grant 617891.

\section*{References}\label{references}
\addcontentsline{toc}{section}{References}

\hypertarget{refs}{}
\hypertarget{ref-baayendavidsonbates2008a}{}
Baayen, R. H., D. J. Davidson, and D. M. Bates. 2008. ``Mixed-Effects
Modeling with Crossed Random Effects for Subjects and Items.''
\emph{Journal of Memory and Language} 59: 390--412.

\hypertarget{ref-bates.maechler.etal:2015}{}
Bates, Douglas, Martin Maechler, Benjamin M. Bolker, and Steven Walker.
2015. ``Fitting Linear Mixed-Effects Models Using lme4.'' \emph{ArXiv},
1406.5823.

\hypertarget{ref-buttonioannidismokrysz2013a}{}
Button, Katherine S, John P A Ioannidis, Claire Mokrysz, Brian A Nosek,
Jonathan Flint, Emma S J Robinson, and Marcus R Munafò. 2013. ``Power
Failure: Why Small Sample Size Undermines the Reliability of
Neuroscience.'' \emph{Nat Rev Neurosci}, Apr.

\hypertarget{ref-clark1973a}{}
Clark, Herbert H. 1973. ``The Language-as-Fixed-Effect Fallacy: A
Critique of Language Statistics in Psychological Research.''
\emph{Journal of Verbal Learning and Verbal Behavior} 12: 335--59.

\hypertarget{ref-cohen1992a}{}
Cohen, Jacob. 1992. ``A Power Primer.'' \emph{Psychological Bulletin}
112 (1): 55--159.

\hypertarget{ref-fox:2016}{}
Fox, John. 2016. \emph{Applied Regression Analysis and Generalized
Linear Models}. 3rd ed. Thousand Oaks, CA: Sage.

\hypertarget{ref-gelman.hill:2006}{}
Gelman, Andrew, and Jennifer Hill. 2006. \emph{Data Analysis Using
Regression and Multilevel/Hierarchical Models}. Cambridge: Cambridge
University Press.

\hypertarget{ref-imai.king.etal:2008jrss}{}
Imai, Kosuke, Gary King, and Elizabeth A. Stuart. 2008.
``Misunderstandings Between Experimentalists and Observationalists About
Causal Inference.'' \emph{Journal of the Royal Statistical Society A}
171, Part 2: 481--502.

\hypertarget{ref-judd.westfall.etal:2012pp}{}
Judd, Charles M., Jacob Westfall, and David A. Kenny. 2012. ``Treating
Stimuli as a Random Factor in Social Psychology: A New and Comprehensive
Solution to a Pervasive but Largely Ignored Problem.'' \emph{J Pers Soc
Psychol} 103 (1): 54--69.

\hypertarget{ref-pinheirobates2000a}{}
Pinheiro, José, and Douglas Bates. 2000. \emph{Mixed-Effects Models in S
and S-PLUS}. Springer New York.

\hypertarget{ref-senn:1994sm}{}
Senn, Stephen. 1994. ``Testing for Baseline Balance in Clinical
Trials.'' \emph{Statistics in Medicine} 13: 1715--26.

\hypertarget{ref-westfall2016}{}
Westfall, Jacob, and Tal Yarkoni. 2016. ``Statistically Controlling for
Confounding Constructs Is Harder Than You Think.'' \emph{PLoS ONE} 3
(11): 1--100.

\section{Methods}\label{methods}

\subsection{Survey}\label{survey}

The analysis was restricted to current volumes. For all articles
published by \emph{B\&L} from 2011 to the 3rd issue of 2013, three
raters (not blinded to the purpose of the experiment) investigated all
published experimental papers (excluding reviews, simulation studies,
editorials etc.). For each experiment reported in a study, the
stimulus/materials sections were investigated for descriptive and
inferential statistics derived from populations that were exhaustively
sampled without error. If a descriptive and/or inferential statistic
(such as mean and standard deviation) were reported, the study was coded
as one where the researchers were interested in a known quantity,
otherwise it was discarded. If an inferential statistic (such as a
\emph{p}-value) was reported, the study was coded as one where
researchers answered that interest with an erroneous parameter estimate,
otherwise as one where researchers did not commit the error. If a
statement of the form that groups were thought equivalent regarding the
parameter was made, such as claims that they were ``matched'', ``equal''
or ``did not differ'', and this statement was backed up by a
\emph{p}-value greater than 0.05, the study was coded as ``accepting the
null''. In cases of rater disagreement, the majority vote was
registered. Representative statements from studies committing an error
are available online at https://github.com/jona-sassenhagen/statfail.

\end{document}